%% file: Phoenixes.tex
\documentclass[final,5p,times,twocolumn,authoryear]{elsarticle}

\usepackage{epsfig}
\usepackage{verbatim}
\usepackage{epstopdf}
\usepackage{graphicx}
\usepackage{aas_macros}
\usepackage{epsf}
\usepackage{amsmath}
\usepackage{xcolor}
\usepackage{tikz}
\usepackage{dblfloatfix}
\usepackage{threeparttable}
\usepackage{enumitem}

\definecolor{darkgreen}{rgb}{0.0,0.5,0.0}
\usepackage[colorlinks,citecolor=darkgreen]{hyperref}

\input{Definitions.tex}
\newcommand{\Mach}{\mathcal{M}}

\makeatletter
\def\ps@pprintTitle{%
  \let\@oddhead\@empty
  \let\@evenhead\@empty
  \def\@oddfoot{\reset@font\hfil\thepage\hfil}
  \let\@evenfoot\@oddfoot
}
\makeatother

\defcitealias{BagchiEtAl98}{B98}
\newcommand{\Bagchi}{{\citetalias{BagchiEtAl98}}}
\defcitealias{SleeEtAl01}{S01}
\newcommand{\Slee}{{\citetalias{SleeEtAl01}}}
\defcitealias{deGasperinEtAl15}{dG15}
\newcommand{\deGasperin}{{\citetalias{deGasperinEtAl15}}}
\defcitealias{MandalEtAl19}{M19}
\newcommand{\Mandal}{{\citetalias{MandalEtAl19}}}
\defcitealias{Schellenberger22}{S22}
\newcommand{\Schellenberger}{{\citetalias{Schellenberger22}}}
\defcitealias{KaleDwarakanath12}{KD12}
\newcommand{\Kale}{{\citetalias{KaleDwarakanath12}}}
\defcitealias{EdlerEtAl22}{E22}
\newcommand{\Edler}{{\citetalias{EdlerEtAl22}}}
\defcitealias{RajaEtAl24}{R24}
\newcommand{\Raja}{{\citetalias{RajaEtAl24}}}
\defcitealias{IchinoheEtAl15}{I15}
\newcommand{\Ichinohe}{{\citetalias{IchinoheEtAl15}}}
\defcitealias{FukazawaEtAl04}{F04}
\newcommand{\Fukazawa}{{\citetalias{FukazawaEtAl04}}}
\defcitealias{SandersonEtAl06}{S06}
\newcommand{\Sanderson}{{\citetalias{SandersonEtAl06}}}
\defcitealias{KaleEtAl18}{K18}
\newcommand{\KaleB}{{\citetalias{KaleEtAl18}}}
\defcitealias{RandallEtAl10}{R10}
\newcommand{\Randall}{{\citetalias{RandallEtAl10}}}
\defcitealias{RahamanEtAl22}{R22}
\newcommand{\Rahaman}{{\citetalias{RahamanEtAl22}}}

\begin{document}

\begin{frontmatter}

\title{Relativistic ions with power-law spectra explain radio phoenixes}

\author{Uri Keshet}
\address{
    Physics Department, Ben-Gurion University of the Negev, POB 653, Be'er-Sheva 84105, Israel; keshet.uri@gmail.com
}

\begin{abstract}
Radio phoenixes are filamentary sources in the intracluster medium (ICM) of galaxy clusters, often extending over $>100$ kpc, arising from fossil radio lobes.
Their soft, curved spectrum is widely attributed to aged relativistic electrons recently accelerated or compressed, but at high frequencies is shown to approach a power-law.
Moreover, the full, curved spectrum is naturally reproduced by secondary $e^{\pm}$ from a pure power-law spectrum of relativistic ions, radiating in highly-magnetized filaments; this model provides a better fit to all phoenixes, with only three free parameters.
Weaker magnetization shifts the curvature to low frequencies, explaining pure power-law phoenixes.
Hadronic high-curvature phoenixes require $e^{\pm}$ heating, by a factor $\gtrsim 15$ if at ICM pressure.
The $\sim$keV Compton- and $\sim$GeV $\pi^0$-decay-peaked counterparts of hadronic phoenixes may be detectable as non-thermal X-rays and $\gamma$-rays.
\end{abstract}

\end{frontmatter}

\section{Introduction}

The intracluster medium (ICM) of galaxy clusters shows diffuse radio emission in various forms, including phoenix-type relics (also known as shocked AGN fossils/ghosts; henceforth phoenixes).
These sources are distinguished from other types of diffuse ICM emission by their soft, and often curved, volume-integrated spectrum, steepening at high frequencies, where the spectral index $\alpha\equiv d\ln F_\nu/d\ln \nu\lesssim-2$; here, $F_\nu$ is the specific flux and $\nu$ is the frequency.
Radio phoenixes are usually highly filamentary, mildly polarized, and associated with fossil relativistic plasma from an active galactic nucleus (AGN) outburst \citep[][and references therein]{vanWeerenEtAl19, RajaEtAl24}.
Unlike their standard (also known as  gischt/flotsam/radio shock) relic counterparts, there is no strong evidence that phoenixes coincide with shocks, although in some cases the phoenix is projected $\gtrsim 200\kpc$ downstream of a possible weak shock \citep{TanakaEtAl10, BotteonEtAl18, Schellenberger22}; some tentative shock associations are not confirmed (\eg \citet{FujitaEtAl04A133, RandallEtAl10} or \citet{BotteonEtAl18, RahamanEtAl22}).

Phoenixes are widely modelled as fossil cosmic-ray electrons (CRE) in radio lobes that previously expanded and cooled, but were recently revived by compression waves induced by a passing shock \citep{EnsslinGopalKrishna01}.
This model and other alternatives, such as recent CRE injection from a compact source, with or without pitch-angle conservation \citep{Kardashev62, JaffePerola73, KomissarovGubanov94}, struggle to explain a large quantity of energetic, rapidly cooling CRE.
In addition to three minimal radio-modelling parameters (typically particle normalization and effective power-law index and magnetic field), such models introduce additional free parameters, in terms of timescales or spectral breaks, and multiple evolutionary stages \citep[\eg flashing or fading;][]{EnsslinGopalKrishna01, KaleDwarakanath12}.
In most cases, such models then fit the observed spectra reasonably well, but imply an ongoing softening at high frequencies.

However, when available at sufficiently high frequencies, the spectral curvature vanishes and $\alpha$ tends to a constant, as illustrated crudely in Fig.~\ref{fig:Alpha} by comparing adjacent-frequency fluxes.
Moreover, as shown below, the full spectrum is well reproduced in all phoenixes by a simple model of only three free-parameters, if CRE are secondaries produced by cosmic-ray ions (CRI), inelastically colliding with ambient nuclei.
Then, there is no need to invoke a recent shock or compact-source injection, as CRI provide a sustained source of energetic CRE.
After outlining the model in \S\ref{sec:Model}, we apply it to phoenix observations in \S\ref{sec:Fit}, and discuss the implications in \S\ref{sec:Discussion}.

\begin{figure}[h]
    \centering
    \includegraphics[width=0.45\textwidth]{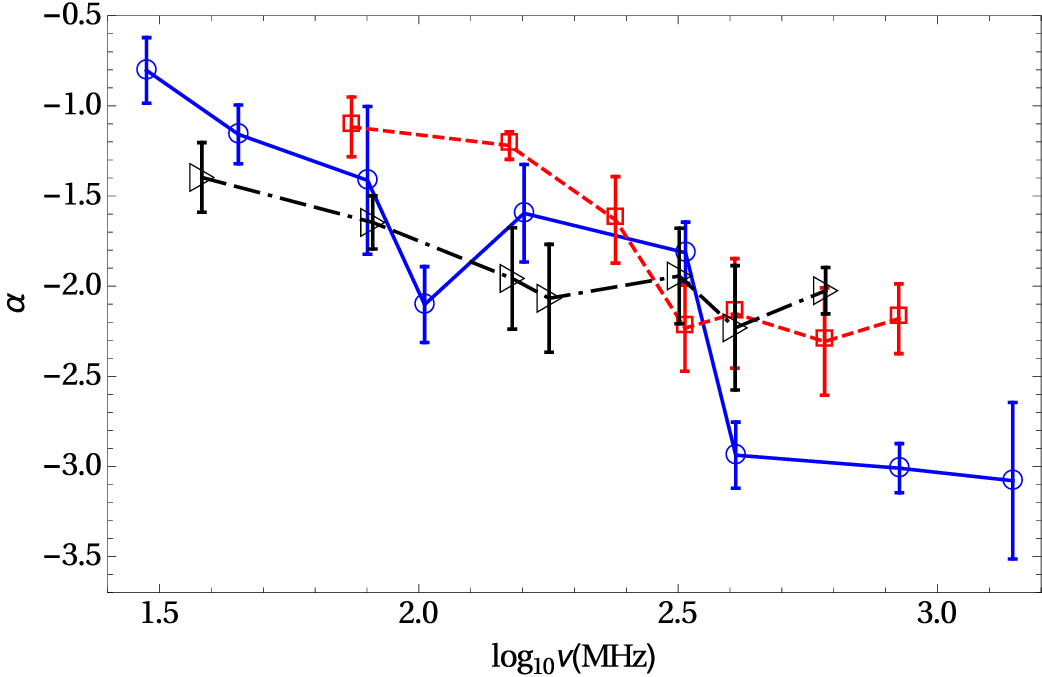}
	\caption{ \label{fig:Alpha}
        Spectral index $\alpha$ as a function of frequency (symbols with lines to guide the eye) for select phoenixes in clusters A85 (blue circles with solid lines), A4038 (red squares, dashed), and A1914 (black triangles, dot-dashed). For data references, see Table \ref{tab:phoenixes}.
        The index $\alpha$ is estimated for each measured frequency by comparing the two adjacent measurements in frequency space.
	}
\end{figure}

We adopt a flat $\Lambda$CDM cosmological model, with a present-day Hubble constant $H_0=70\km\se^{-1}\Mpc^{-1}$, an $\Omega_m=0.3$ mass fraction,
an $f_b=0.17$ baryon fraction, and a $76\%$ hydrogen mass fraction, yielding a mean particle mass $\bar{m}\simeq 0.59m_p$.
Error bars indicate $68.3\%$ containment, projected onto one parameter in figures or within multivariate fits for model parameters.

\section{Hadronic phoenix model}
\label{sec:Model}

Consider the injected CRE gyrating in strong transverse magnetic fields $B$ in some part of the phoenix.
The resulting specific synchrotron luminosity is approximately \citep[\eg][]{Keshet24}
\begin{equation}\label{eq:Synch}
L_\nu(\nu) \simeq m_e^2 c^4 \left(\frac{\nu}{3\nu_B}\right)^{1/2}\,N\left(E=\alpha_E h\sqrt{\nu_B\nu} \right) \coma
\end{equation}
where $\nu$ is the emitted photon frequency, $N(E)$ is the CRE number per unit energy $E$ in the (filamentary) radiating volume $V$,
\begin{equation}\label{eq:nuB}
\nu_B\equiv \frac{8 m_e c^3}{9\sigma_T e B} \coma
\end{equation}
$\alpha_E$ is the electromagnetic coupling constant, $m_e$ and $e$ are the electron mass and charge, $h$ is the Planck constant, $\sigma_T$ is the Thomson cross section, and $c$ is the speed of light.

The CRE are shown to cool rapidly, so their evolution approximately follows the steady-state injection--loss equation
\begin{equation}\label{eq:EnergyEq}
0 \simeq \partial_t N(E) = \dot{N}_+(E/f) + \psi\partial_E(E^2 N) \coma
\end{equation}
where $\dot{N}_+$ is the differential CRE injection rate,
\begin{equation}\label{eq:Cooling}
\psi \simeq \frac{B^2 \sigma_T }{6\pi m_e^2 c^3}
\end{equation}
is the synchrotron-dominated cooling coefficient,
and we allowed for fast post-injection CRE heating by some factor $f>1$.
The steady-state CRE distribution is therefore 
\begin{equation}\label{eq:nuB}
N(E) = \frac{f}{\psi E^2} \int_{E/f}^\infty \dot{N}_+(E')dE' \fin
\end{equation}

For $N_i(E_i)$ CRI per unit energy in $V$, and a number density $n$ of ambient nuclei, the injection rate is
\begin{equation}\label{eq:Injection}
  \dot{N}_+(E) = c n \int N_i(E_i) \frac{d\sigma_{e}(E_i,E)}{d E} \, dE_i \coma
\end{equation}
where we assumed homogeneity in $V$; otherwise, $N_i$ and $n$ should be replaced by weighted averages maintaining Eq.~\eqref{eq:Injection}.
The differential inclusive cross-section
$d\sigma_e(E_i,E)/dE$ for $e^{\pm}$ production is computed using the formulation of \citet[][with corrected parameters and cutoffs; T. Kamae \& H. Lee, private communications 2010]{KamaeEtAl06}.
Finally, we parameterize $N_i(E_i)=C_{\mbox{\scriptsize{TeV}}} (E_i/\mbox{TeV})^{-p}$, where the normalization $C_{\mbox{\scriptsize{TeV}}}$ and spectral index $p$ are constants.

The observed specific flux, integrated over the entire phoenix, can now be modelled as
\begin{equation}\label{eq:Flux}
  F_\nu(\nu) = \frac{(1+z)L_\nu[(1+z)\nu]}{4\pi d_L^2} \coma
\end{equation}
where $d_L(z)$ is the luminosity distance to the cluster, and $z$ is its redshift.
The $F_\nu(\nu)$ profile, measured at multiple frequencies, thus becomes a function of only three free parameters:
the CRI spectral index $p$, the combination $f^2 B$ regulating the $E(\nu)$ dependence, and an overall normalization factor which can be chosen as $A=f^4n\, C_{\mbox{\scriptsize{TeV}}}$.

\section{Hadronic model applied to phoenix observations}
\label{sec:Fit}

\begin{figure}[h]
    \centering
    \includegraphics[width=0.5\textwidth]{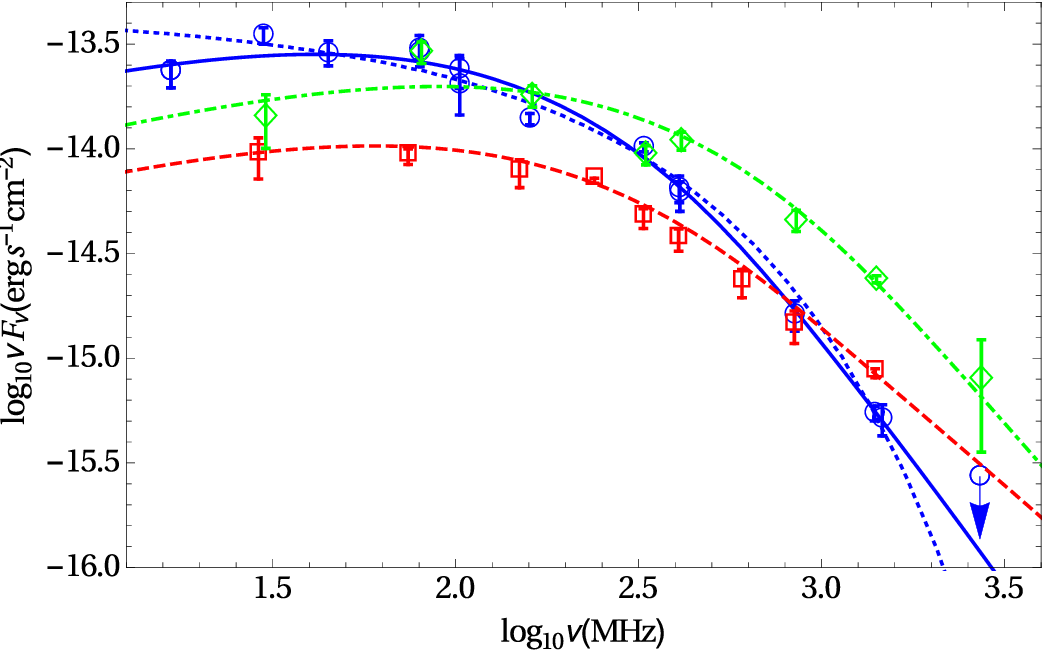}
	\caption{ \label{fig:CurvedPhoenixes}
        Measured (symbols) and best-fit (curves) high-curvature spectra of radio phoenixes in clusters A85 (blue circles and solid curve), A133 (green diamonds, dot-dashed), and A4038 (red squares, dashed).
        The CRE compression model \citep{EnsslinGopalKrishna01} is demonstrated for A85 (dotted blue).
        For data references, see Table \ref{tab:phoenixes}.
	}
\end{figure}

\begin{figure}[h]
    \centering
    \includegraphics[width=0.5\textwidth]{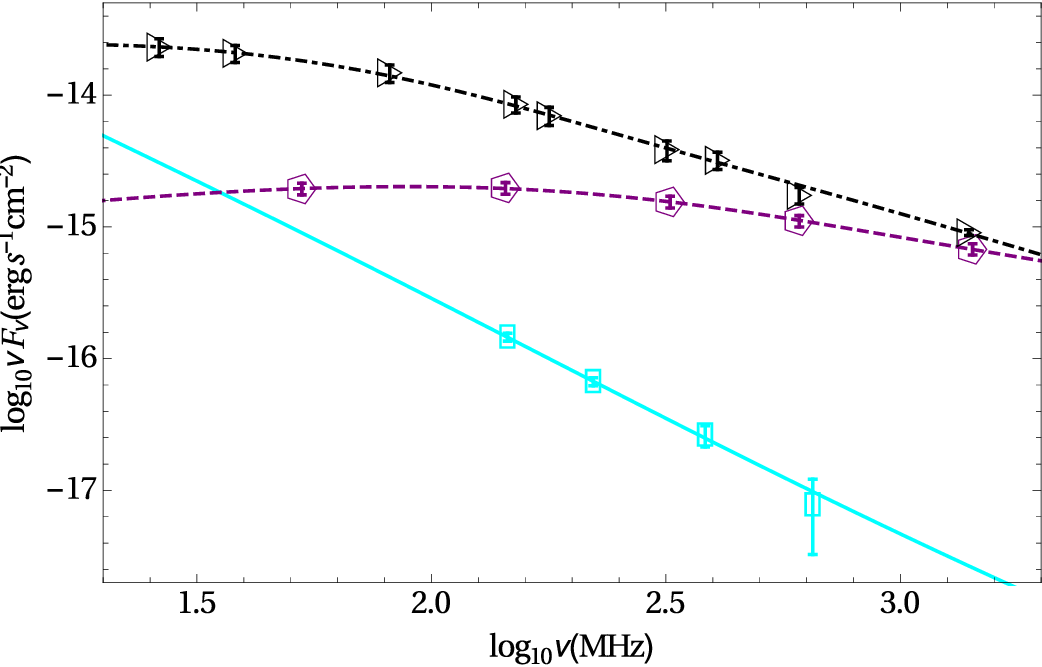}
	\caption{ \label{fig:NonCurvedPhoenixes}
        Same as Fig.~\ref{fig:CurvedPhoenixes}, but for phoenixes with little to no spectral curvature, in clusters A1033 (purple pentagons, dashed curve), A1914 (black triangles, dot-dashed), and A2108 (cyan rectangles, solid).
        See Table \ref{tab:phoenixes} for references.
	}
\end{figure}

The properties of six phoenixes with published integrated radio spectra and X-ray profiles are summarized in Table \ref{tab:phoenixes}.
Three of these phoenixes, in clusters A85, A133, and A4038, show strong curvature; their spectrum and best-fit hadronic models are shown in Fig.~\ref{fig:CurvedPhoenixes}.
The other three phoenixes, in clusters A1033, A1914, and A2108, show little to no curvature; they are presented in Fig.~\ref{fig:NonCurvedPhoenixes}.
The table provides the best-fit model parameters and the chi-squared per degree of freedom, $\chi^2/\mathfrak{n}$.

\begin{table*}[h!]
\begin{threeparttable}
{\small %
    	\caption{
            \label{tab:phoenixes}
    		Phoenix parameters and modelling. \\
            \textbf{Columns:} (1) Cluster name; (2) Redshift $z$ (from the SIMBAD$^1$ database); (3) CRI spectral index $p$; (4) CRE $E(\nu)$ parameter $f^2B$ ($\mu$G units); (5) Logarithm of $A=f^4 n\, C_{\mbox{\scriptsize{TeV}}}$ (cm$^{-3}\erg^{-1}$ units) normalization;
            (6) Chi-squared per degree of freedom; (7) Approximate radius $r$ of the phoenix, defined with respect to the X-ray brightness peak of the cluster (kpc units); (8) Approximate phoenix length $D$ (longest axis; kpc units); (9) ICM pressure based on X-rays (eV cm$^{-3}$ units); (10) Mach number $\Mach$ inferred from $p$; (11) Minimal value $f_{min}=\langle\sin^2\alpha_p \rangle^{1/4} f$, corresponding to full, $B^2/8\pi=\langle\sin^2\alpha_p \rangle P_{\mbox{\tiny{ICM}}}$ magnetization; (12) Radio references$^2$ for $F_\nu$ and $D$; (13) X-ray references$^2$ for $r$ and $P_{\mbox{\tiny{ICM}}}$.
        }
{\small
        \begin{tabular}{lc|cccc|ccc|cc|ll}
    		Cluster &  $z$ & $p$ & $f^2B$ & $\mbox{log}_{10}A$ & $\chi^2/\mathfrak{n}$ & $r$ & $D$ & $P_{\mbox{\tiny{ICM}}}$ & $\Mach$ & $f_{min}$ & Radio Ref. & X-ray Ref. \\
    		(1) & (2) & (3) & (4) & (5) & (6) & (7) & (8) & (9) & (10) & (11) & (12) & (13) \\
    		\hline
            A85 & 0.0556 & $6.22\pm0.55$ & $625\pm79$ & $41.0\pm1.8$ & $3.05$ & 450 & 330 & 41 & $1.40\pm0.05$ & 15 & \Bagchi, \Raja & \Slee, \Ichinohe \\
            A133 & 0.0565 & $5.59\pm0.28$ & $1306\pm161$ & $42.9\pm0.9$ & $4.31$ & 33 & 70 & 101 & $1.45\pm0.04$ & 20 & \Slee & \Randall \\
            A1033 & 0.1259 & $3.11\pm0.41$ & $948\pm563$ & $50.2\pm1.2$ & $0.01$ & 75 & 360 & 96 & $2.14_{-0.24}^{+0.44}$ & 15 & \Edler & \deGasperin \\
            A1914 & 0.1660 & $3.81\pm0.17$ & $206\pm92$ & $49.5\pm0.6$ & $0.31$ & 130 & 510 & 98 & $1.79\pm0.07$ & 3 & \Mandal & \Mandal, \Rahaman \\
            A2108 & 0.0898 & $5.08\pm0.94$ & $9\pm144$ & $45.7\pm10.9$ & $0.23$ & 100 & 140 & 50 & $1.52\pm0.17$ & 0.2 & \Schellenberger & \Schellenberger \\
            A4038 & 0.0297 & $4.69\pm0.48$ & $795\pm215$ & $44.8\pm1.5$ & $1.83$ & 30 & 55 & 108 & $1.58\pm0.10$ & 12 & \Kale & \KaleB, \Sanderson, \Fukazawa \\
        \end{tabular}
}
\begin{tablenotes}
    \item[1] {\small \href{http://simbad.u-strasbg.fr/simbad/}{http://simbad.u-strasbg.fr/simbad/}}
    \item[2] {\small
    Reference abbreviations:
    {\Bagchi} -- \citet{BagchiEtAl98};
    {\deGasperin} -- \citet{deGasperinEtAl15};
    {\Edler} -- \citet{EdlerEtAl22};
    {\Fukazawa} -- \citet{FukazawaEtAl04};
    {\Ichinohe} -- \citet{IchinoheEtAl15};
    {\KaleB} -- \citet{KaleEtAl18};
    {\Kale} -- \citet{KaleDwarakanath12};
    {\Mandal} -- \citet{MandalEtAl19};
    {\Randall} -- \citet{RandallEtAl10};
    {\Rahaman} -- \citet{RahamanEtAl22};
    {\Raja} -- \citet{RajaEtAl24};
    {\Sanderson} -- \citet{SandersonEtAl06};
    {\Slee} -- \citet{SleeEtAl01}.
    {\Schellenberger} -- \citet{Schellenberger22};
    }
    \item[3] {\small In A1033 and A2108, we adopt a typical $n=0.006\cm^{-3}$ nucleon density, as no published deprojected densities are available.}
\end{tablenotes}
}
\end{threeparttable}
\end{table*}

In all six phoenixes, and in other, less explored phoenixes \citep[not shown due to missing redshift, X-ray deprojection, etc.; \eg][]{vanWeerenEtAl09}, the model provides a very good fit to the data, considering the inherent difficulties in calibrating different experiments and consistently integrating the low surface brightness over the entire phoenix in each band.
Indeed, the large, $\chi^2/\mathfrak{n}>2$ values obtained in A85 and A133 arise from large deviations between measurements in adjacent bands.
Note that while the fit for the phoenix in A2108 is very good and $p$ is well determined, the other parameters are highly uncertain due to its negligible spectral curvature.

The hadronic model works better than previous alternatives, despite having fewer free parameters.
For illustration, Fig.~\ref{fig:CurvedPhoenixes} shows, for A85, also the CRE compression model \citep{EnsslinGopalKrishna01}, underproducing the curvature at low frequencies and overproducing it at large frequencies.
In phoenixes of limited curvature, alternative models provide fits \citep[\eg][]{vanWeerenEtAl09} as good as the hadronic model, but with at least two additional free parameters.
In addition, these fits are not always natural or self-consistent, requiring for example very recent, $\lesssim 10\Myr$ compact-source injection, or a total age insufficient for inflating an average-size phoenix.

The results suggest that each phoenix is powered by CRI of a pure, soft power-law energy spectrum, of index $p>3$.
Such power-laws are usually attributed to diffusive shock acceleration in a weak shock, which in this case could have taken place long ago, as CRI energy is effectively conserved.
A weak ICM shock cannot penetrate the high-pressure, relativistic plasma of a radio lobe, but could shock its low-pressure regions.
Alternatively, the CRI may have been accelerated in weak shocks at lobe boundaries, their bow shocks, or the dissipation of a relativistic compact-source outflows.
Adopting the $p=2(\Mach^2+1)/(\Mach^2-1)$ relation \citep[\eg][]{Krymskii77} for non-relativistic upstream gas of $\Gamma=5/3$ adiabatic index,
the implied shocks are all weak, with Mach numbers $\Mach<2.5$, or even $\Mach<2$ aside from $\Mach\simeq 2.1$ in A1033; see Table \ref{tab:phoenixes}.

ICM shocks typically strengthen with increasing radius $r$, but we find no correlation between the $\Mach$ and $r$ values of the phoenixes; the Spearman ranked correlation coefficient $\rho\simeq -0.26$ indicates an insignificant anti-correlation.
The Mach numbers also do not correlate with reported shocks, usually far from the phoenix, in A1033 \citep{deGasperinEtAl15}, A1914 \citep{BotteonEtAl18}, and A2108 \citep{Schellenberger22}.
Putative shocks at radio-lobe edges are generally expected to grow weaker with increasing size, which we approximate by the length of their longest axis, $D$.
Again, we find no anti-correlation between $\Mach$ and $D$; $\rho\simeq+0.43$ indicates an insignificant positive correlation.

The parameter $f^2B$ is found to be large, approaching $1\mbox{ mG}$, and even exceeding it in A133.
For such values, the energy density in the filamentary magnetic field greatly exceeds that of the CRE, and is comparable to the total pressure, justifying some CRE heating, \eg by magnetic reconnection, as anticipated in Eq.~\eqref{eq:EnergyEq}, and the neglect of Compton losses in Eq.~\eqref{eq:Cooling}.
Assuming that the radiating volume is not over-pressurized with respect to the surrounding ICM pressure $P_{\mbox{\tiny{ICM}}}$, we may require $B^2/8\pi<\langle\sin^2\alpha_p\rangle P_{\mbox{\tiny{ICM}}}$, where $\alpha_p$ is the CRE pitch angle and $\langle\ldots\rangle$ averages over the CRE distribution.
X-ray observations that gauge the ambient pressure then impose a lower limit $f_{min}$ on $\langle\sin^2\alpha_p\rangle^{1/4}f$.
We find $f_{min}$ values close to the mean (energy-weighted) parent proton-to-electron energy ratio, $\langle E_i/E \rangle$, reaching $\gtrsim15$ in high-curvature phoenixes, but cannot explain the heating mechanism given the unknown plasma conditions.
Observable electron energies span a factor $<10$ in each phoenix, so our energy-independent $f$ may be justified.

The $A=f^4n\, C_{\mbox{\scriptsize{TeV}}}$ normalization varies strongly among phoenixes, spanning $\sim$nine orders of magnitude in Table \ref{tab:phoenixes}.
This dispersion is only slightly reduced when accounting for the relatively small differences in phoenix volumes and $f_{min}$ limits.
Magnetic depletion is expected to significantly diminish $n$ with respect to its ICM value, but this effect cannot govern the dispersion in $A$.
Rather, given the large values of $p$, selection effects require very large $A$ when the spectrum is soft.
Indeed, $A$ and $p$ are strongly anti-correlated ($\rho=-0.94$; $p$-value $0.0048$).

Even under extreme assumptions --- a single, narrow filament with $V=(1\kpc)^2D$ and strong, $n=0.01n_{\mbox{\tiny{ICM}}}$ plasma depletion --- most phoenixes require a CRI energy density less than the $u_i(>10\GeV)\simeq 10^{-13.6}\erg\cm^{-3}$ inferred in the ICM outside phoenixes from radio \citep{Keshet10} and \gama-ray \citep{Keshet25PaperII} observations.
The best fits in A1914 and A2108 require more energy in this limit, but even for $V=(1\kpc)D^2\ll D^3$, A1914 requires a plausible $u_i=10^{-12.7\pm0.6}\erg\cm^{-3}$; in A2108, the $A$ uncertainty is prohibitively large.

\section{Discussion}
\label{sec:Discussion}

A simple hadronic model (see \S\ref{sec:Model}) explains all available radio-phoenix spectra, as demonstrated in \S\ref{sec:Fit} (Figs.~\ref{fig:CurvedPhoenixes} and \ref{fig:NonCurvedPhoenixes}), with only three free parameters (columns 3--5 in Table \ref{tab:phoenixes}).
The model naturally accounts for the full spectrum of each phoenix, including its curvature and high-frequency power-law (Fig.~\ref{fig:Alpha}), without the additional parameters introduced by leptonic models, and without invoking a recent violent event to energize the CRE.
Hadronic phoenixes at the ICM pressure require strong magnetization, and some CRE heating by a factor $f\gtrsim \langle E_i/E\rangle$, reaching $f\gtrsim15$ when the curvature manifests at high synchrotron frequencies (compare rows 4, 9, and 11 in Table \ref{tab:phoenixes}).

For simplicity, we used a one-zone model, although some spectral variations are expected across a phoenix.
In particular, more peripheral regions should present a softer spectrum with more curvature, due to a combination of weaker $B$ and some CRE reaching the magnetized regions after partial Compton cooling.
Indeed, spectral variations reported across some phoenixes \citep{KaleEtAl18, RajaEtAl24} are qualitatively consistent with these trends.
CRE rapidly lose all their energy in the strong $B$, and their diffusion is quenched, so diffusion--cooling spectral corrections \citep{Keshet24} are negligible.

The strong magnetization inferred likely constitutes a considerable fraction of the pressure, thus locally depleting the plasma; the filamentary structure probably traces these strong magnetic fields.
CRI are unlikely to be confined to these fields for long, so secondary CRE injection should persist also, or mostly, outside the strongly magnetized regions.
Consequently, the Compton and $\pi^0\to\gamma\gamma$ counterparts of the synchrotron emission should produce coincident, non-filamentary, non-thermal X-ray and \gama-ray signals, which would test the model and break its parameter degeneracies.
Protons of few GeV energy produce $\sim$keV photons, as their secondary (non-heated) CRE Compton-scatter cosmic microwave background (CMB) photons, and $\sim$GeV photons, through $\pi^0$ decay.
At higher X-ray and \gama-ray energies, it would be easier to observe the harder phoenixes, like in A1033 where $p\simeq 3.1$.

Non-phoenix forms of diffuse radio emission from the ICM, including minihalos, giant halos, mega-halos, and even standard relics, can all be accounted for by the same cluster-wide CRI distribution that is approximately flat, spectrally and spatially, as anticipated from radio observations \citep{Keshet10} and recently \gama-ray-detected in Coma \citep[][and references therein]{Keshet24, KushnirEtAl24} and in stacked clusters \citep{Keshet25PaperII}.
These CRI permeate the ICM by means of strong diffusion, recently measured in halos and behind standard relics, also responsible for their high-frequency and peripheral softening \citep{Keshet24}.
However, radio lobes are likely shielded from these cluster CRI by strong magnetic fields, explaining why we find different CRI populations powering phoenixes.

Like phoenixes, standard, shock-driven relics too are widely explained by invoking leptonic models that are unnatural and inconsistent with observations \citep{Keshet10,Keshet24} --- for example, their spectra are too different from shock-acceleration predictions \citep[][and references therein]{vanWeerenEtAl19}, too universal, and too uniform along the shock front --- but are fully accounted for by the cluster-wide CRI distribution, if CRI are delayed in a shell behind the shock before rapidly diffusing away \citep{Keshet24}.
Thus, standard relics and phoenixes are distinguished by their different CRI populations, as well as by the apparent absence of a shock in the latter.

\paragraph*{Acknowledgements} \begin{small}
I thank Yuri Lyubarsky for helpful discussions.
This research was supported by the Israel Science Foundation (ISF grant No. 2126/22), and used the SIMBAD database, operated at CDS, Strasbourg, France.
\end{small}

\bibliographystyle{elsarticle-harv-author-truncate}
\bibliography{Virial}

\end{document}

%% file: Definitions.tex
\newcommand{\eg}{\emph{e.g.,} }

\newcommand{\be}{\begin{equation}}
\newcommand{\ee}{\end{equation}}
\newcommand{\bea}{\begin{equation*}}
\newcommand{\eea}{\end{equation*}}
\newcommand{\beqr}{\begin{eqnarray} \nonumber}
\newcommand{\eeqr}{\end{eqnarray}}
\newcommand{\beqrb}{\begin{eqnarray}}
\newcommand{\eeqrb}{\nonumber \end{eqnarray}}
\newcommand{\fin}{\mbox{ .}}
\newcommand{\coma}{\mbox{ ,}}

\newcommand{\cm}{\mbox{ cm}}

\newcommand{\se}{\mbox{ s}}

\newcommand{\Myr}{\mbox{ Myr}}
\newcommand{\erg}{\mbox{ erg}}

\newcommand{\km}{\mbox{ km}}

\newcommand{\kpc}{\mbox{ kpc}}
\newcommand{\Mpc}{\mbox{ Mpc}}

\newcommand{\GeV}{\mbox{ GeV}}

\newcommand{\gama}{$\gamma$}

\newcommand{\mynewcommand}[2]{\ifdefined #1 \else \newcommand{#1}{#2} \fi}

\mynewcommand{\apj}{ApJ}     
\mynewcommand{\apjl}{ApJL}     
\mynewcommand{\apjs}{ApJS}    
\mynewcommand{\aap}{A\&A}    
\mynewcommand{\nat}{Nature}  

%% file: Phoenixes.bbl
\begin{thebibliography}{29}
\expandafter\ifx\csname natexlab\endcsname\relax\def\natexlab#1{#1}\fi
\providecommand{\url}[1]{\texttt{#1}}
\providecommand{\href}[2]{#2}
\providecommand{\path}[1]{#1}
\providecommand{\DOIprefix}{doi:}
\providecommand{\ArXivprefix}{arXiv:}
\providecommand{\URLprefix}{URL: }
\providecommand{\Pubmedprefix}{pmid:}
\providecommand{\doi}[1]{\href{http://dx.doi.org/#1}{\path{#1}}}
\providecommand{\Pubmed}[1]{\href{pmid:#1}{\path{#1}}}
\providecommand{\bibinfo}[2]{#2}
\ifx\xfnm\relax \def\xfnm[#1]{\unskip,\space#1}\fi
\bibitem[{{Bagchi} et~al.(1998){Bagchi}, {Pislar} and {Lima
  Neto}}]{BagchiEtAl98}
\bibinfo{author}{{Bagchi}, J.}, \bibinfo{author}{{Pislar}, V.},
  \bibinfo{author}{{Lima Neto}, G.B.}, \bibinfo{year}{1998}.
\newblock \bibinfo{title}{{The diffuse, relic radio source in Abell 85:
  estimation of cluster-scale magnetic field from inverse Compton X-rays}}.
\newblock \bibinfo{journal}{\mnras} \bibinfo{volume}{296},
  \bibinfo{pages}{L23--l28}.
\newblock \DOIprefix\doi{10.1046/j.1365-8711.1998.01589.x},
  \href{http://arxiv.org/abs/astro-ph/9803020}{{\tt arXiv:astro-ph/9803020}}.
\bibitem[{{Botteon} et~al.(2018){Botteon}, {Gastaldello} and
  {Brunetti}}]{BotteonEtAl18}
\bibinfo{author}{{Botteon}, A.}, \bibinfo{author}{{Gastaldello}, F.},
  \bibinfo{author}{{Brunetti}, G.}, \bibinfo{year}{2018}.
\newblock \bibinfo{title}{{Shocks and cold fronts in merging and massive galaxy
  clusters: new detections with Chandra}}.
\newblock \bibinfo{journal}{\mnras} \bibinfo{volume}{476},
  \bibinfo{pages}{5591--5620}.
\newblock \DOIprefix\doi{10.1093/mnras/sty598},
  \href{http://arxiv.org/abs/1707.07038}{{\tt arXiv:1707.07038}}.
\bibitem[{{de Gasperin} et~al.(2015){de Gasperin}, {Ogrean}, {van Weeren},
  {Dawson}, {Br{\"u}ggen}, {Bonafede} and {Simionescu}}]{deGasperinEtAl15}
\bibinfo{author}{{de Gasperin}, F.}, \bibinfo{author}{{Ogrean}, G.A.},
  \bibinfo{author}{{van Weeren}, R.J.}, \bibinfo{year}{2015}.
\newblock \bibinfo{title}{{Abell 1033: birth of a radio phoenix}}.
\newblock \bibinfo{journal}{\mnras} \bibinfo{volume}{448},
  \bibinfo{pages}{2197--2209}.
\newblock \DOIprefix\doi{10.1093/mnras/stv129},
  \href{http://arxiv.org/abs/1501.00043}{{\tt arXiv:1501.00043}}.
\bibitem[{{Edler} et~al.(2022){Edler}, {de Gasperin}, {Brunetti}, {Botteon},
  {Cuciti}, {van Weeren}, {Cassano}, {Shimwell}, {Br{\"u}ggen} and
  {Drabent}}]{EdlerEtAl22}
\bibinfo{author}{{Edler}, H.W.}, \bibinfo{author}{{de Gasperin}, F.},
  \bibinfo{author}{{Brunetti}, G.}, \bibinfo{year}{2022}.
\newblock \bibinfo{title}{{Abell 1033: Radio halo and gently reenergized tail
  at 54 MHz}}.
\newblock \bibinfo{journal}{\aap} \bibinfo{volume}{666}, \bibinfo{pages}{A3}.
\newblock \DOIprefix\doi{10.1051/0004-6361/202243737},
  \href{http://arxiv.org/abs/2207.11040}{{\tt arXiv:2207.11040}}.
\bibitem[{{En{\ss}lin} and {Gopal-Krishna}(2001)}]{EnsslinGopalKrishna01}
\bibinfo{author}{{En{\ss}lin}, T.A.}, \bibinfo{author}{{Gopal-Krishna}},
  \bibinfo{year}{2001}.
\newblock \bibinfo{title}{Reviving fossil radio plasma in clusters of galaxies
  by adiabatic compression in environmental shock waves}.
\newblock \bibinfo{journal}{\aap} \bibinfo{volume}{366},
  \bibinfo{pages}{26--34}.
\newblock \DOIprefix\doi{10.1051/0004-6361:20000198},
  \href{http://arxiv.org/abs/arXiv:astro-ph/0011123}{{\tt
  arXiv:arXiv:astro-ph/0011123}}.
\bibitem[{{Fujita} et~al.(2004){Fujita}, {Sarazin}, {Reiprich}, {Andernach},
  {Ehle}, {Murgia}, {Rudnick} and {Slee}}]{FujitaEtAl04A133}
\bibinfo{author}{{Fujita}, Y.}, \bibinfo{author}{{Sarazin}, C.L.},
  \bibinfo{author}{{Reiprich}, T.H.}, \bibinfo{year}{2004}.
\newblock \bibinfo{title}{{XMM-Newton Observations of A133: A Weak Shock
  Passing through the Cool Core}}.
\newblock \bibinfo{journal}{\apj} \bibinfo{volume}{616},
  \bibinfo{pages}{157--168}.
\newblock \DOIprefix\doi{10.1086/424807},
  \href{http://arxiv.org/abs/astro-ph/0407596}{{\tt arXiv:astro-ph/0407596}}.
\bibitem[{{Fukazawa} et~al.(2004){Fukazawa}, {Makishima} and
  {Ohashi}}]{FukazawaEtAl04}
\bibinfo{author}{{Fukazawa}, Y.}, \bibinfo{author}{{Makishima}, K.},
  \bibinfo{author}{{Ohashi}, T.}, \bibinfo{year}{2004}.
\newblock \bibinfo{title}{Asca compilation of x-ray properties of hot gas in
  elliptical galaxies and galaxy clusters: Two breaks in the temperature
  dependences}.
\newblock \bibinfo{journal}{\pasj} \bibinfo{volume}{56},
  \bibinfo{pages}{965--1009}.
\newblock \href{http://arxiv.org/abs/arXiv:astro-ph/0411745}{{\tt
  arXiv:arXiv:astro-ph/0411745}}.
\bibitem[{{Ichinohe} et~al.(2015){Ichinohe}, {Werner}, {Simionescu}, {Allen},
  {Canning}, {Ehlert}, {Mernier} and {Takahashi}}]{IchinoheEtAl15}
\bibinfo{author}{{Ichinohe}, Y.}, \bibinfo{author}{{Werner}, N.},
  \bibinfo{author}{{Simionescu}, A.}, \bibinfo{year}{2015}.
\newblock \bibinfo{title}{{The growth of the galaxy cluster Abell 85: mergers,
  shocks, stripping and seeding of clumping}}.
\newblock \bibinfo{journal}{\mnras} \bibinfo{volume}{448},
  \bibinfo{pages}{2971--2986}.
\newblock \DOIprefix\doi{10.1093/mnras/stv217},
  \href{http://arxiv.org/abs/1410.1955}{{\tt arXiv:1410.1955}}.
\bibitem[{{Jaffe} and {Perola}(1973)}]{JaffePerola73}
\bibinfo{author}{{Jaffe}, W.J.}, \bibinfo{author}{{Perola}, G.C.},
  \bibinfo{year}{1973}.
\newblock \bibinfo{title}{{Dynamical Models of Tailed Radio Sources in Clusters
  of Galaxies}}.
\newblock \bibinfo{journal}{\aap} \bibinfo{volume}{26}, \bibinfo{pages}{423}.
\bibitem[{{Kale} and {Dwarakanath}(2012)}]{KaleDwarakanath12}
\bibinfo{author}{{Kale}, R.}, \bibinfo{author}{{Dwarakanath}, K.S.},
  \bibinfo{year}{2012}.
\newblock \bibinfo{title}{{Multi-frequency Studies of Radio Relics in the
  Galaxy Clusters A4038, A1664, and A786}}.
\newblock \bibinfo{journal}{\apj} \bibinfo{volume}{744}, \bibinfo{pages}{46}.
\newblock \DOIprefix\doi{10.1088/0004-637X/744/1/46},
  \href{http://arxiv.org/abs/1109.3261}{{\tt arXiv:1109.3261}}.
\bibitem[{{Kale} et~al.(2018){Kale}, {Parekh} and {Dwarakanath}}]{KaleEtAl18}
\bibinfo{author}{{Kale}, R.}, \bibinfo{author}{{Parekh}, V.},
  \bibinfo{author}{{Dwarakanath}, K.S.}, \bibinfo{year}{2018}.
\newblock \bibinfo{title}{{A study of spectral curvature in the radio relic in
  Abell 4038 using the uGMRT}}.
\newblock \bibinfo{journal}{\mnras} \bibinfo{volume}{480},
  \bibinfo{pages}{5352--5361}.
\newblock \DOIprefix\doi{10.1093/mnras/sty2227},
  \href{http://arxiv.org/abs/1808.04057}{{\tt arXiv:1808.04057}}.
\bibitem[{{Kamae} et~al.(2006){Kamae}, {Karlsson}, {Mizuno}, {Abe} and
  {Koi}}]{KamaeEtAl06}
\bibinfo{author}{{Kamae}, T.}, \bibinfo{author}{{Karlsson}, N.},
  \bibinfo{author}{{Mizuno}, T.}, \bibinfo{year}{2006}.
\newblock \bibinfo{title}{{Parameterization of {$\gamma$}, $e^{+/-}$, and
  Neutrino Spectra Produced by p-p Interaction in Astronomical Environments}}.
\newblock \bibinfo{journal}{\apj} \bibinfo{volume}{647},
  \bibinfo{pages}{692--708}.
\newblock \DOIprefix\doi{10.1086/505189},
  \href{http://arxiv.org/abs/arXiv:astro-ph/0605581}{{\tt
  arXiv:arXiv:astro-ph/0605581}}.
\bibitem[{{Kardashev}(1962)}]{Kardashev62}
\bibinfo{author}{{Kardashev}, N.S.}, \bibinfo{year}{1962}.
\newblock \bibinfo{title}{{Nonstationarity of Spectra of Young Sources of
  Nonthermal Radio Emission}}.
\newblock \bibinfo{journal}{\sovast} \bibinfo{volume}{6}, \bibinfo{pages}{317}.
\bibitem[{{Keshet}(2010)}]{Keshet10}
\bibinfo{author}{{Keshet}, U.}, \bibinfo{year}{2010}.
\newblock \bibinfo{title}{Common origin for radio relics and halos: galaxy
  cluster-wide, homogeneous cosmic-ray distribution, and evolving magnetic
  fields}.
\newblock \bibinfo{journal}{ArXiv e-prints}
  \DOIprefix\doi{10.48550/arxiv.1011.0729},
  \href{http://arxiv.org/abs/1011.0729}{{\tt arXiv:1011.0729}}.
\bibitem[{{Keshet}(2024)}]{Keshet24}
\bibinfo{author}{{Keshet}, U.}, \bibinfo{year}{2024}.
\newblock \bibinfo{title}{Radio haloes and relics from extended cosmic-ray ion
  distributions with strong diffusion in galaxy clusters}.
\newblock \bibinfo{journal}{\mnras} \bibinfo{volume}{527},
  \bibinfo{pages}{1194--1215}.
\newblock \DOIprefix\doi{10.1093/mnras/stad3154},
  \href{http://arxiv.org/abs/2303.08146}{{\tt arXiv:2303.08146}}.
\bibitem[{{Keshet}(2025)}]{Keshet25PaperII}
\bibinfo{author}{{Keshet}, U.}, \bibinfo{year}{2025}.
\newblock \bibinfo{title}{{Galaxy-cluster-stacked Fermi-LAT II: extended
  central hadronic signal}}.
\newblock \bibinfo{journal}{arXiv e-prints} ,
  \bibinfo{pages}{arXiv:2502.19494}\DOIprefix\doi{10.48550/arXiv.2502.19494},
  \href{http://arxiv.org/abs/2502.19494}{{\tt arXiv:2502.19494}}.
\bibitem[{{Komissarov} and {Gubanov}(1994)}]{KomissarovGubanov94}
\bibinfo{author}{{Komissarov}, S.S.}, \bibinfo{author}{{Gubanov}, A.G.},
  \bibinfo{year}{1994}.
\newblock \bibinfo{title}{Relic radio galaxies: evolution of synchrotron
  spectrum}.
\newblock \bibinfo{journal}{\aap} \bibinfo{volume}{285},
  \bibinfo{pages}{27--43}.
\bibitem[{{Krymskii}(1977)}]{Krymskii77}
\bibinfo{author}{{Krymskii}, G.F.}, \bibinfo{year}{1977}.
\newblock \bibinfo{title}{A regular mechanism for the acceleration of charged
  particles on the front of a shock wave}.
\newblock \bibinfo{journal}{Akademiia Nauk SSSR Doklady} \bibinfo{volume}{234},
  \bibinfo{pages}{1306--1308}.
\bibitem[{{Kushnir} et~al.(2024){Kushnir}, {Keshet} and
  {Waxman}}]{KushnirEtAl24}
\bibinfo{author}{{Kushnir}, D.}, \bibinfo{author}{{Keshet}, U.},
  \bibinfo{author}{{Waxman}, E.}, \bibinfo{year}{2024}.
\newblock \bibinfo{title}{{Coma cluster $\gamma$-ray and radio emission is
  consistent with a secondary electron origin for the radio halo}}.
\newblock \bibinfo{journal}{arXiv e-prints} ,
  \bibinfo{pages}{arXiv:2404.13111}\DOIprefix\doi{10.48550/arXiv.2404.13111},
  \href{http://arxiv.org/abs/2404.13111}{{\tt arXiv:2404.13111}}.
\bibitem[{{Mandal} et~al.(2019){Mandal}, {Intema}, {Shimwell}, {van Weeren},
  {Botteon}, {R{\"o}ttgering}, {Hoang}, {Brunetti}, {de Gasperin},
  {Giacintucci}, {Hoekstra}, {Stroe}, {Br{\"u}ggen}, {Cassano}, {Shulevski},
  {Drabent} and {Rafferty}}]{MandalEtAl19}
\bibinfo{author}{{Mandal}, S.}, \bibinfo{author}{{Intema}, H.T.},
  \bibinfo{author}{{Shimwell}, T.W.}, \bibinfo{year}{2019}.
\newblock \bibinfo{title}{{Ultra-steep spectrum emission in the merging galaxy
  cluster Abell 1914}}.
\newblock \bibinfo{journal}{\aap} \bibinfo{volume}{622}, \bibinfo{pages}{A22}.
\newblock \DOIprefix\doi{10.1051/0004-6361/201833992},
  \href{http://arxiv.org/abs/1811.08430}{{\tt arXiv:1811.08430}}.
\bibitem[{{Rahaman} et~al.(2022){Rahaman}, {Raja} and {Datta}}]{RahamanEtAl22}
\bibinfo{author}{{Rahaman}, M.}, \bibinfo{author}{{Raja}, R.},
  \bibinfo{author}{{Datta}, A.}, \bibinfo{year}{2022}.
\newblock \bibinfo{title}{{On the detection of multiple shock fronts in A1914
  using deep Chandra X-ray observations}}.
\newblock \bibinfo{journal}{\mnras} \bibinfo{volume}{509},
  \bibinfo{pages}{5821--5835}.
\newblock \DOIprefix\doi{10.1093/mnras/stab3115},
  \href{http://arxiv.org/abs/2110.12297}{{\tt arXiv:2110.12297}}.
\bibitem[{{Raja} et~al.(2024){Raja}, {Rahaman}, {Datta} and
  {Smirnov}}]{RajaEtAl24}
\bibinfo{author}{{Raja}, R.}, \bibinfo{author}{{Rahaman}, M.},
  \bibinfo{author}{{Datta}, A.}, \bibinfo{author}{{Smirnov}, O.M.},
  \bibinfo{year}{2024}.
\newblock \bibinfo{title}{{A Multifrequency View of the Radio Phoenix in the
  A85 Cluster}}.
\newblock \bibinfo{journal}{\apj} \bibinfo{volume}{975}, \bibinfo{pages}{125}.
\newblock \DOIprefix\doi{10.3847/1538-4357/ad7585},
  \href{http://arxiv.org/abs/2309.14244}{{\tt arXiv:2309.14244}}.
\bibitem[{{Randall} et~al.(2010){Randall}, {Clarke}, {Nulsen}, {Owers},
  {Sarazin}, {Forman} and {Murray}}]{RandallEtAl10}
\bibinfo{author}{{Randall}, S.W.}, \bibinfo{author}{{Clarke}, T.E.},
  \bibinfo{author}{{Nulsen}, P.E.J.}, \bibinfo{year}{2010}.
\newblock \bibinfo{title}{Radio and deep chandra observations of the disturbed
  cool core cluster abell 133}.
\newblock \bibinfo{journal}{\apj} \bibinfo{volume}{722},
  \bibinfo{pages}{825--846}.
\newblock \DOIprefix\doi{10.1088/0004-637X/722/1/825},
  \href{http://arxiv.org/abs/1008.2921}{{\tt arXiv:1008.2921}}.
\bibitem[{{Sanderson} et~al.(2006){Sanderson}, {Ponman} and
  {O'Sullivan}}]{SandersonEtAl06}
\bibinfo{author}{{Sanderson}, A.J.R.}, \bibinfo{author}{{Ponman}, T.J.},
  \bibinfo{author}{{O'Sullivan}, E.}, \bibinfo{year}{2006}.
\newblock \bibinfo{title}{A statistically selected chandra sample of 20 galaxy
  clusters - i. temperature and cooling time profiles}.
\newblock \bibinfo{journal}{\mnras} \bibinfo{volume}{372},
  \bibinfo{pages}{1496--1508}.
\newblock \DOIprefix\doi{10.1111/j.1365-2966.2006.10956.x},
  \href{http://arxiv.org/abs/arXiv:astro-ph/0608423}{{\tt
  arXiv:arXiv:astro-ph/0608423}}.
\bibitem[{{Schellenberger} et~al.(2022){Schellenberger}, {Giacintucci},
  {Lovisari}, {O'Sullivan}, {Vrtilek}, {David}, {Melin}, {Lal}, {Ettori},
  {Kolokythas}, {Sereno} and {Raychaudhury}}]{Schellenberger22}
\bibinfo{author}{{Schellenberger}, G.}, \bibinfo{author}{{Giacintucci}, S.},
  \bibinfo{author}{{Lovisari}, L.}, \bibinfo{year}{2022}.
\newblock \bibinfo{title}{{The Unusually Weak and Exceptionally Steep Radio
  Relic in A2108}}.
\newblock \bibinfo{journal}{\apj} \bibinfo{volume}{925}, \bibinfo{pages}{91}.
\newblock \DOIprefix\doi{10.3847/1538-4357/ac3b5a},
  \href{http://arxiv.org/abs/2111.09225}{{\tt arXiv:2111.09225}}.
\bibitem[{{Slee} et~al.(2001){Slee}, {Roy}, {Murgia}, {Andernach} and
  {Ehle}}]{SleeEtAl01}
\bibinfo{author}{{Slee}, O.B.}, \bibinfo{author}{{Roy}, A.L.},
  \bibinfo{author}{{Murgia}, M.}, \bibinfo{year}{2001}.
\newblock \bibinfo{title}{Four extreme relic radio sources in clusters of
  galaxies}.
\newblock \bibinfo{journal}{\aj} \bibinfo{volume}{122},
  \bibinfo{pages}{1172--1193}.
\newblock \DOIprefix\doi{10.1086/322105},
  \href{http://arxiv.org/abs/arXiv:astro-ph/0105267}{{\tt
  arXiv:arXiv:astro-ph/0105267}}.
\bibitem[{{Tanaka} et~al.(2010){Tanaka}, {Furuzawa}, {Miyoshi}, {Tamura} and
  {Takata}}]{TanakaEtAl10}
\bibinfo{author}{{Tanaka}, N.}, \bibinfo{author}{{Furuzawa}, A.},
  \bibinfo{author}{{Miyoshi}, S.J.}, \bibinfo{year}{2010}.
\newblock \bibinfo{title}{{Suzaku Observations of the Merging Cluster Abell 85:
  Temperature Map and Impact Direction}}.
\newblock \bibinfo{journal}{\pasj} \bibinfo{volume}{62}, \bibinfo{pages}{743}.
\newblock \DOIprefix\doi{10.1093/pasj/62.3.743},
  \href{http://arxiv.org/abs/1006.4328}{{\tt arXiv:1006.4328}}.
\bibitem[{{van Weeren} et~al.(2019){van Weeren}, {de Gasperin}, {Akamatsu},
  {Br{\"u}ggen}, {Feretti}, {Kang}, {Stroe} and {Zandanel}}]{vanWeerenEtAl19}
\bibinfo{author}{{van Weeren}, R.J.}, \bibinfo{author}{{de Gasperin}, F.},
  \bibinfo{author}{{Akamatsu}, H.}, \bibinfo{year}{2019}.
\newblock \bibinfo{title}{Diffuse radio emission from galaxy clusters}.
\newblock \bibinfo{journal}{\ssr} \bibinfo{volume}{215}, \bibinfo{pages}{16}.
\newblock \DOIprefix\doi{10.1007/s11214-019-0584-z},
  \href{http://arxiv.org/abs/1901.04496}{{\tt arXiv:1901.04496}}.
\bibitem[{{van Weeren} et~al.(2009){van Weeren}, {R{\"o}ttgering},
  {Br{\"u}ggen} and {Cohen}}]{vanWeerenEtAl09}
\bibinfo{author}{{van Weeren}, R.J.}, \bibinfo{author}{{R{\"o}ttgering},
  H.J.A.}, \bibinfo{author}{{Br{\"u}ggen}, M.}, \bibinfo{author}{{Cohen}, A.},
  \bibinfo{year}{2009}.
\newblock \bibinfo{title}{A search for steep spectrum radio relics and halos
  with the gmrt}.
\newblock \bibinfo{journal}{\aap} \bibinfo{volume}{508},
  \bibinfo{pages}{75--92}.
\newblock \DOIprefix\doi{10.1051/0004-6361/200912501},
  \href{http://arxiv.org/abs/0910.2970}{{\tt arXiv:0910.2970}}.

\end{thebibliography}
